\documentclass[prl,twocolumn,showpacs,nofootinbib,superscriptaddress]{revtex4-1}

\usepackage{amsfonts}
\usepackage{amsmath}
\usepackage{amssymb}
\usepackage{bm}
\usepackage{dcolumn}
\usepackage{epsfig}
\usepackage{graphicx}
\usepackage{graphics}
\usepackage[latin1]{inputenc}
\usepackage{latexsym}
\usepackage{rotating}
\usepackage{xspace} % Sensible space treatment at end of simple macros
\usepackage[usenames]{color}
\usepackage{hyperref}

\usepackage{ulem}
\definecolor {darkgreen}{rgb}{0.2,0.7,0.2}

\definecolor{purple}{rgb}{0.5,0,0.5}

\newcommand\be{\begin{equation}}
\newcommand\ba{\begin{eqnarray}}
\newcommand\ee{\end{equation}}
\newcommand\ea{\end{eqnarray}}

\newcommand{\nn}{\nonumber}

\newcommand{\Newt}{{\mbox{\tiny Newt}}}

\newcommand{\NS}{*}

\begin{document}
\title{Strong Binary Pulsar Constraints \\ on Lorentz Violation in Gravity} 

\author{Kent Yagi}
\affiliation{Department of Physics, Montana State University, Bozeman, MT 59717, USA.}

\author{Diego Blas}
\affiliation{CERN, Theory Division, 1211 Geneva, Switzerland.}

\author{Nicol\'as Yunes}
\affiliation{Department of Physics, Montana State University, Bozeman, MT 59717, USA.}

\author{Enrico Barausse}
\affiliation{CNRS, UMR 7095, Institut d'Astrophysique de Paris, 98bis Bd Arago, 75014 Paris, France}
\affiliation{Sorbonne Universit\'es, UPMC Univ Paris 06, UMR 7095, 98bis Bd Arago, 75014 Paris, France}

\date{\today}

%%%%%%%%%%%%%%%%%%%%%%%%%%%%%%%%%%%%%%%%%%%%%%%%%
\begin{abstract} 

Binary pulsars are excellent laboratories to test the building blocks of Einstein's theory of General Relativity. One of these is Lorentz symmetry which states that physical phenomena appear the same for all inertially moving observers. We study the effect of violations of Lorentz symmetry in the orbital evolution of binary pulsars and find that it induces a much more rapid decay of the binary's orbital period due to the emission of dipolar radiation. The absence of such behaviour in recent observations allows us to place the most stringent constraints on Lorentz violation in gravity, thus verifying one of the cornerstones of Einstein's theory much more accurately than any previous gravitational observation.    

\end{abstract}


%04.30.Db Wave generation and sources
% 04.50.Kd Modified theories of gravity
% 04.25.-g Approximation methods; equations of motion
%04.25.Nx Post-Newtonian approximation; perturbation theory; related approximations

\onecolumngrid
\begin{flushright}
\small CERN-PH-TH/2013-167
\end{flushright}
\vspace{-.5cm}
\twocolumngrid

\maketitle

%Word limit is 3500 excluding title, abstract, ref and acknowledgments.
% A figure corresponds to 340 words.
%An equation corresponds to 16 words per row.

%%%%%%%%%%%%%%%%%%%%%%%%%%%%%%%%%%%
{\emph{Introduction.}}--- Lorentz symmetry allowed physicists to reconcile Maxwell's electromagnetism with the principle of relativity, which states that the outcome of experiments should be the same for all inertial observers. This reconciliation was the basis of Einstein's theory of Special Relativity. Einstein later formulated General Relativity (GR) as a Lorentz symmetric completion of Newtonian gravity. Today, much of theoretical physics is built on Lorentz symmetry. In particular, it is a cornerstone of the standard model of particle physics. Given how embedded this symmetry is in our understanding of Nature, any observation of its violation would shake theoretical physics at its core.

The experimental verification of Lorentz symmetry has a long history. Today, particle physics experiments 
constrain Lorentz violation in the standard model to an exquisite degree~\cite{Kostelecky:2008ts}. But 
the same is not yet true for gravitational phenomena. Solar System~\cite{will-living,Blas:2011zd,Jacobson:2008aj}, certain binary pulsar~\cite{Bell:1995jz,Shao:2013wga,Jacobson:2008aj}, and cosmological~\cite{Zuntz:2008zz,Audren:2013dwa,Jacobson:2008aj} observations have been used to derive bounds on Lorentz violation in gravity, but those are either weaker or partial, focusing on preferred-frame effects. Lorentz symmetry has not yet been tested in regimes where gravity is strong and the gravitational interaction is  non-linear, such as mergers of neutron stars (NSs) and black holes collisions.   

One may wonder about the necessity to test Lorentz symmetry in gravity, given the tight constraints coming from particle physics. In fact, if the Lorentz violating effects in gravity were to percolate into particle physics in a manner consistent with these constraints, such effects would be unobservable in gravitational experiments. However, this is not necessarily the case; different mechanisms have been put forth leading to Lorentz violating effects in gravity that can be much larger than those in particle physics, see e.g. Ref.~\cite{Liberati:2013xla} or the discussion in Ref.~\cite{long-paper} for a review.

%%%%%%%%%%%%%%%%%%%%%%%%%%%%%%%%%%%
{\emph{Lorentz Violation in Gravity.}}--- Let us consider modified theories that violate Lorentz symmetry due to the existence of a preferred time direction at each point of space and time. This establishes a local preferred frame that violates Lorentz symmetry because the velocity of inertial observers with respect to this frame is in principle observable. Much work has gone into searching for this preferred time direction in high energy physics \cite{Kostelecky:2008ts,Liberati:2013xla}, but here we will concentrate on constraining low-energy Lorentz violation in the gravity sector. As explained in detail and shown in~\cite{Jacobson:2000xp,Jacobson:2010mx}, one can model this preferred frame without loss of generality through a timelike unit vector, the so-called \AE ther field $U^{\mu}$. The dynamics of this vector field can be prescribed by two related theories: Einstein-\AE ther theory~\cite{Jacobson:2000xp} and khronometric theory \cite{Blas:2010hb}. In the former, the vector field is generic, while in the latter it is related to the existence of a global preferred time coordinate~\cite{Jacobson:2010mx,Blas:2010hb}. Both theories can be considered as low-energy descriptions of high-energy completions of GR responsible for Lorentz violation~\cite{Jacobson:2000xp}. In the khronometric case, a possible completion is Ho\v rava gravity \cite{Horava:2009uw,Blas:2009qj}, which hinges on Lorentz violation to yield a power-counting renormalizable completion of GR.

The couplings of the \AE ther field to gravity are directly related to Lorentz violation. In the low-energy limit, these are controlled by two sets of constants that affect different phenomena. One of them, parameterized by $(\alpha_1,\alpha_2)$, controls preferred-frame effects in the weak-field and has been constrained by Solar System~\cite{will-living,Foster:2005dk,Bailey:2006fd,Blas:2011zd} and pulsar observations~\cite{Bell:1995jz,Shao:2013wga}. The other set, parametrized by $(c_{+},c_{-})$ in Einstein-\AE ther theory and $(\beta,\lambda)$ in khronometric theory, controls other, more relativistic, Lorentz-violating effects. Theoretical constraints on this set follow from stability considerations and the absence of gravitational Cherenkov radiation~\cite{Jacobson:2008aj,Blas:2010hb,Donnelly:2010qd}. Cosmological observations can also be used to constrain this set of parameters in the khronometric case~\cite{Audren:2013dwa, Blas:2010hb}, but not efficiently in the Einstein-\AE ther theory~\cite{Jacobson:2008aj,Zuntz:2008zz}. Likewise, as  we show in~\cite{long-paper}, constraints on the strong-field versions of $(\alpha_1,\alpha_2)$~\cite{Shao:2012eg,Shao:2013wga} do not efficiently bound $(c_+,c_-)$ or $(\lambda,\beta)$. In contrast, Lorentz-violation in the dissipative sector of strongly-gravitating systems may lead to large observable effects \cite{Foster:2007gr}. 
Here we calculate these effects rigorously for the first time and place constraints on both sets of parameters.

%%%%%%%%%%%%%%%%%%%%%%%%%%%%%%%%%%%
{\emph{Binary Pulsars as Probes of Lorentz Violation.}}--- Binary pulsars consist of a NS in orbit around either another NS or a less compact companion, like a white dwarf. NSs are strongly-gravitating  sources because their masses and radii are $M_{\NS} \in (1,2.4) M_{\odot}$ and $R_{\NS} \in (10,15) \; {\rm{km}}$, leading to gravitational fields and compactnesses $C_{\NS} = -\Phi_{\Newt}/c^{2} = G M_{\NS}/(R_{\NS} c^{2}) \in (0.1,0.3)$ , where $G$ is Newton's constant, $c$ the speed of light and $\Phi_{\Newt}$ is the Newtonian gravitational potential at the surface of the star.

Lorentz-violating theories induce corrections to the orbital evolution of binary pulsars. In GR, the orbital period decreases due to the emission of gravitational waves, which occurs in a quadrupolar fashion because the component's masses and the center of mass vector are conserved. In Lorentz-violating theories, however, the orbital period decays much more rapidly because of the existence of dipolar radiation. Such radiation is present because the center of gravitational mass does not necessarily coincide with the center of inertial mass. This results in a time-varying dipole moment that emits radiation as the objects spiral into each other, dramatically accelerating the rate of orbital decay.

%%%%%%%%%%%%%%%%%%%%%%%%%%%%%%%%%%%
{\emph{Orbital Period Decay.}}--- Let us consider a binary in a circular orbit with component masses $m_1$ and $m_{2}$. The orbit-averaged rate of orbital decay in a post-Newtonian (PN) expansion~\footnote{A PN expansion is one in the ratio of the orbital velocity to the speed of light. Henceforth, we set $c = 1$, except to label PN orders.} is~\cite{Foster:2007gr}
\be
\frac{\dot{P}_{b}}{P_{b}} = - \frac{192 \pi}{5} \left(\frac{2 \pi G m}{P_{b}}\right)^{5/3} \left(\frac{m_1 m_2}{m^2}\right) \frac{1}{P_{b}} \left<\mathcal{A}\right>\,,
\label{flux-AE}
\ee
in Einstein-\AE ther theory, $m = m_{1} + m_{2}$ is the total mass, $P_{b}$ is the orbital period and we have defined
\begin{align}
\label{A-AE}
\left<\mathcal{A}\right> &= \left[\left(1 - s_{1}\right) \left(1 - s_{2}\right)\right]^{2/3}  
\left( \mathcal{A}_1 + \mathcal{S} \mathcal{A}_2 + \mathcal{S}^{2} \mathcal{A}_3  \right) 
\nn \\
 &+ \frac{5}{32} \left(s_1 - s_2\right)^2 \mathcal{A}_{4}  \left( \frac{P_b}{2 \pi G m} \right)^{2/3} + {\cal{O}} \left( \frac{1}{c^{2}} \right)\,.
\end{align}
In this equation, the quantities $s_{A}$ are sensitivity parameters that will be defined in the next section, ${\cal{S}} = m_{1} s_{2}/m + m_{2} s_{1}/m$ is a mass-weighted sensitivity average and $(\mathcal{A}_{1},\mathcal{A}_{2},\mathcal{A}_{3},\mathcal{A}_{4})$ are certain functions of the coupling constants $(c_{+},c_{-})$~\footnote{This expression was first derived in~\cite{Foster:2007gr}, but with mistakes that we correct in Eqs.~\eqref{flux-AE} and~\eqref{A-AE}, as described in~\cite{long-paper}.}~\cite{long-paper}. An identical expression is obtained in khronometric gravity, except that the functions $(\mathcal{A}_{1},\mathcal{A}_{2},\mathcal{A}_{3},\mathcal{A}_{4})$ now depend on the coupling constants $(\beta,\lambda)$ of that theory~\cite{long-paper}. Equation~(\ref{flux-AE}) reduces to the GR result when $\mathcal{A}=1$, while the second term in Eq.~(\ref{A-AE}) corresponds to dipolar radiation.

For systems that are widely separated, like all observed binary pulsars, the orbital decay rate is dominated by the term with the least powers of $G m/P_{b}$. This is because $G m/P_{b} = {\cal{O}}(10^{-10})$ for a typical NS binary with a 1-hour orbital period. Clearly then, the dipole term dominates the orbital decay rate for all Lorentz-violating theories, unless $s_{1} - s_{2} \approx 0$.

%%%%%%%%%%%%%%%%%%%%%%%%%%%%%%%%%%%
{\emph{Neutron Star Sensitivities.}}--- As is clear from Eqs.~(\ref{flux-AE}) and~(\ref{A-AE}), the orbital decay rate in Lorentz-violating theories depends on the sensitivity parameters $s_{A}$. These quantities are a measure of how the binding energy of a star changes as a function of its relative motion with respect to the preferred frame.

The sensitivities can only be computed once a moving NS star solution in Lorentz-violating theories is obtained; we will work in a slow-motion approximation to first order in the velocity $v \ll 1$, which is sufficient for their calculation without loss of generality~\cite{Foster:2006az,Foster:2007gr,long-paper}. We begin by constructing the most general metric and \AE ther field ansatz for a slowly-moving NS. At ${\cal{O}}(v^{0})$, this ansatz contains only 2 free functions of the radial coordinate. At ${\cal{O}}(v^{1})$, an appropriate gauge choice reduces the ansatz to 3 (2) additional  functions of the radial coordinate and polar angle in Einstein-\AE ther (khronometric) theory.

With this metric ansatz, one can expand the  field equations in small velocity and solve them numerically order by order. To ${\cal{O}}(v^{0})$, one obtains the Lorentz-violating version of the Tolman-Oppenheimer-Volkoff equation, which describes the NS structure and leads to the NS mass-radius relation. To ${\cal{O}}(v^{1})$, one finds a system of partial differential equations that can be decoupled into ordinary differential equations with tensor spherical harmonics. This decoupling is similar to what occurs with the Einstein equations for generic metric perturbations about a Schwarzschild black hole background~\cite{Regge:1957rw}. We establish this  result here for the first time in Lorentz-violating theories, which is crucial to easily find a numerical solution. The ${\cal{O}}(v^{1})$ equations prescribe the behavior of the metric  and \AE ther perturbations, which in turn determine the sensitivities.

When numerically solving the differential equations of structure one must close them by choosing an equation of state (EoS). This equation fixes the pressure as a function of the energy density in the NS interior. We restrict attention to spherically-symmetric, non-rotating, cold (and thus old) NSs, as these are appropriate simplifications for binary pulsar studies~\cite{bhat,2.01NS}. For these stars, there are several realistic EoSs available; we here explore four representative examples: APR~\cite{APR}, SLy~\cite{SLy}, Shen~\cite{Shen1,Shen2} and Lattimer-Swesty with nuclear incompressibility of $220 \; {\rm{MeV}}$ (LS220)~\cite{LS}, the last two with temperature $0.1 \; {\rm{MeV}}$, neutrino-less and in $\beta$-equilibrium. Since Lorentz violations in the matter sector are strongly constrained
experimentally, viable Lorentz-violating modifications to the EoSs are
forced to be small and would not produce any detectable effect on
the systems we are considering. Thus, we can focus, without loss of
generality, on violations in the gravity sector alone (see e.g.
Ref.~\cite{Liberati:2013xla} and the discussion in
Ref.~\cite{long-paper} for a review of possible mechanisms yielding
violations of Lorentz symmetry in gravity that do not percolate into
the matter sector).

All numerical solutions are obtained as follows. In the NS interior, we numerically integrate the differential equations from some core radius $r_{\rm c} \ll R_{\NS}$ to the NS surface $R_{\NS}$, defined as the radius at which the internal pressure vanishes. We then use the value of the interior solution at the NS surface as initial conditions to numerically integrate the exterior equations from the surface to some matching radius $r_{\rm m}\gg R_{\NS}$. The exterior solution is then compared to an approximate analytic solution, calculated asymptotically as an expansion about spatial infinity. This comparison allows us to read out the mass of the NS and to guarantee that the metric is continuous and first-order differentiable at the matching surface. All throughout we use a fourth-order Runge-Kutta algorithm for numerical integrations, checking that our results are robust to changes in discretization, size of core radius and location of matching surface. 

\begin{figure}[ht]
\begin{center}
\begin{tabular}{l r}
\includegraphics[width=8.7cm,clip=true, bb=0 0 792 612]{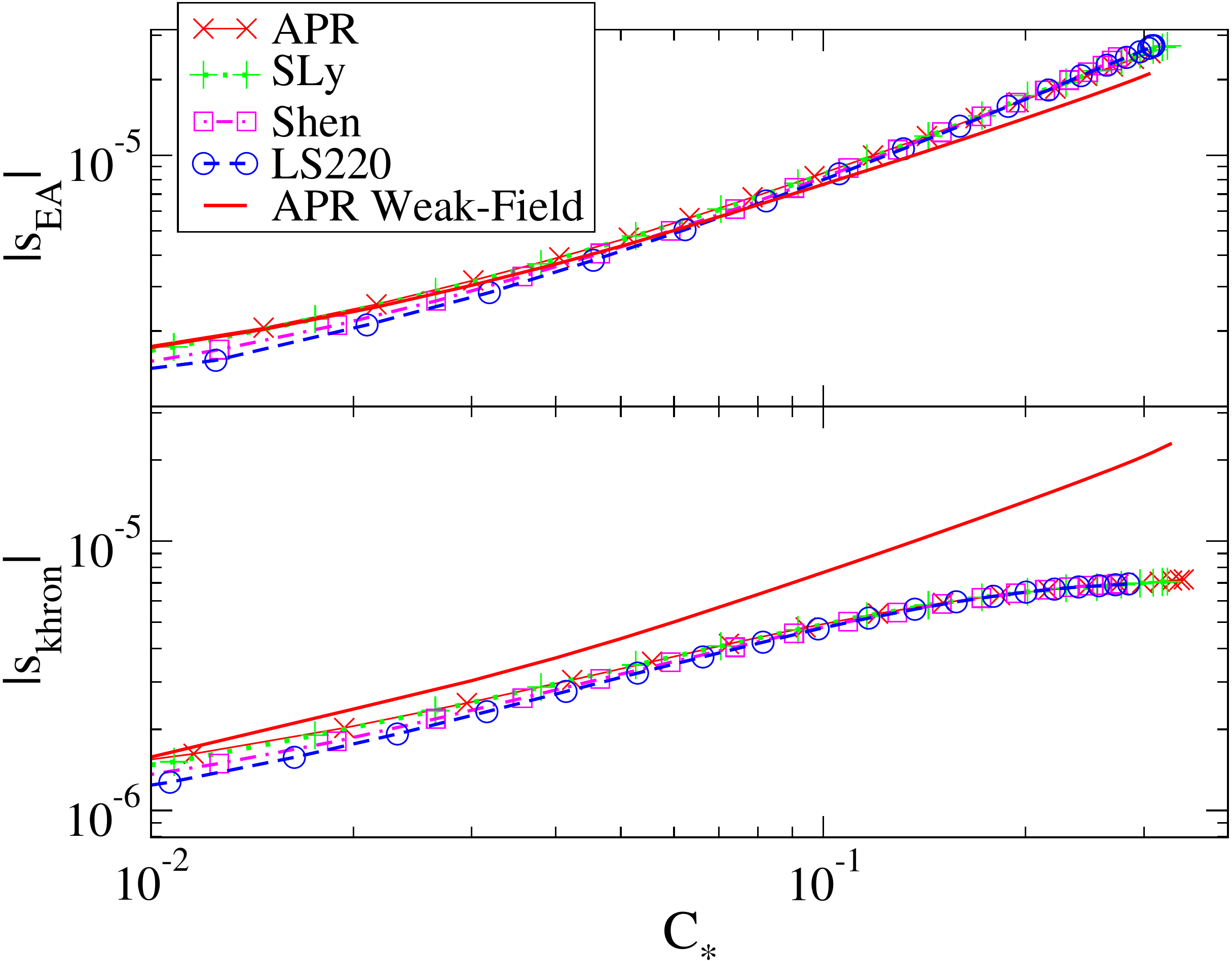}  
\end{tabular}
\caption{\label{fig:sensitivities}
 NS sensitivities in Lorentz-violating theories. We plot the absolute value of $s_A$ in Einstein-\AE ther (top) and khronometric theory (bottom) as a function of the NS compactness for $(\alpha_{1},\alpha_{2})$ that saturate Solar System constraints. We use the constraints from Solar System tests and not from binary pulsars because the latter constrain not only $(\alpha_{1},\alpha_{2})$ but also $c_{+}$ and $c_{-}$~\cite{long-paper}. We choose $c_+$, $c_{-}$ and $\beta$ equal to $10^{-4}$, with $\lambda$ determined by $\beta$ for the chosen values of $(\alpha_1,\alpha_{2})$. Different curves correspond to different EoS. Observe that $s_A$ increases with increasing $C_{\NS}$. The solid (red) curve is the low-compactness approximation to $s_A$~\cite{Foster:2007gr}, which disagrees with our results for realistic NS compactness ($C_{\NS} \sim 0.1$).}
\end{center}
\end{figure}

Figure~\ref{fig:sensitivities} shows $s_{A}$ in Einstein-\AE ther theory (top panel) and khronometric theory (bottom panel) as a function of the NS compactness. 
The sensitivities can be approximated in the low-compactness regime as a linear function in the ratio of the binding energy to the NS mass~\cite{Foster:2007gr}. This expression for $s_A$ is shown as a solid (red) line in Fig.~\ref{fig:sensitivities}. Observe that this approximation is inaccurate for NSs since $C_{\NS} \in (0.1,0.3)$ approximately, and  the ratio of the binding energy to the mass is not such a small number. We find that the low-compactness approximation underestimates the correct value of the NS sensitivity by as much as $200\%$ for realistic NS compactnesses.   

\begin{figure*}[ht]
\begin{center}
\begin{tabular}{l r}
\includegraphics[width=7.5cm,clip=true,bb=0 0 360 358]{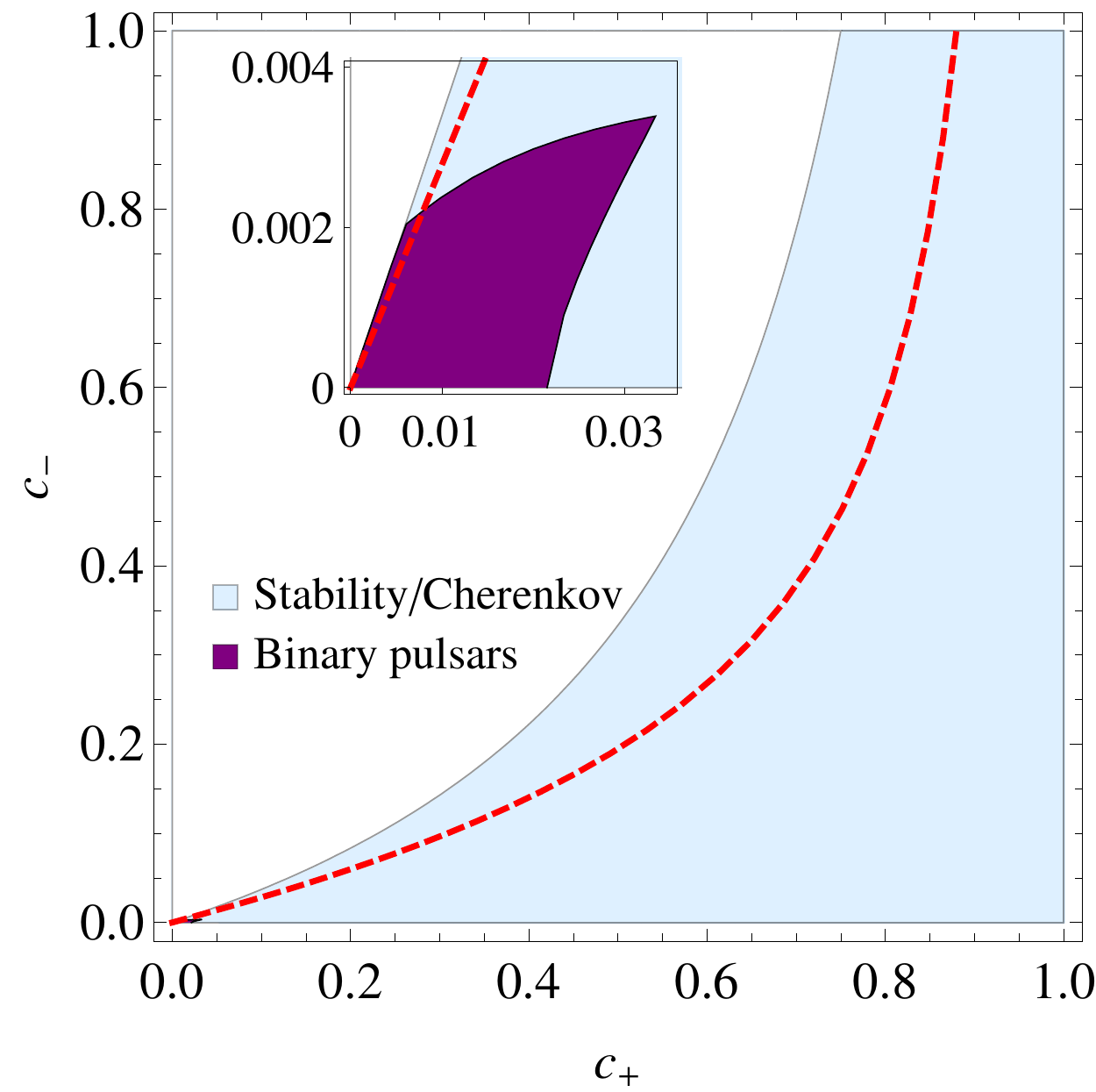}  &
\includegraphics[width=7.7cm,clip=true,bb=0 0 360 350]{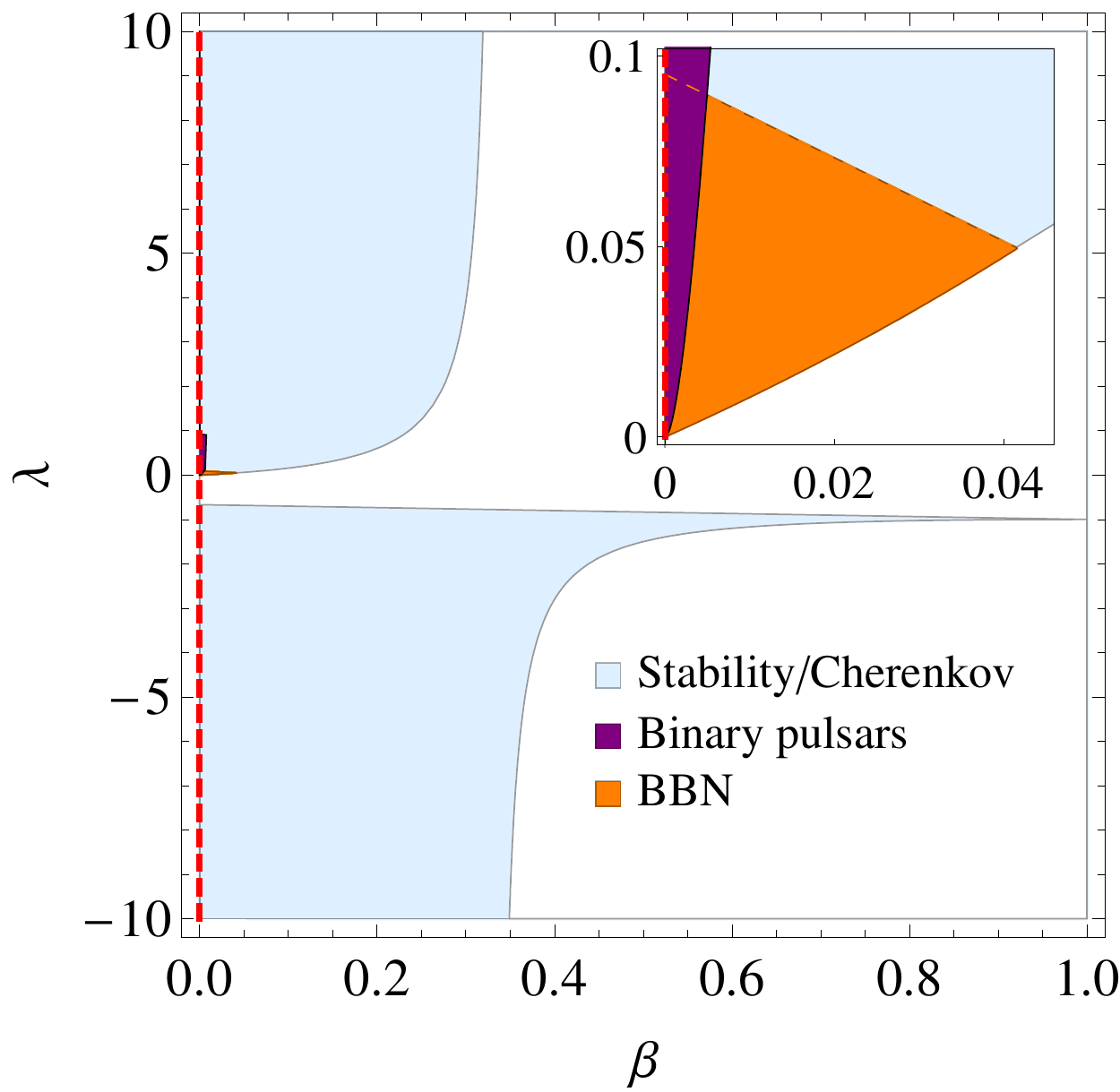}  
\end{tabular}
\caption{\label{fig:constraints}
 Binary pulsar constraints on Lorentz-violating theories. The dark, purple shaded surfaces are allowed regions in the 2-dimensional coupling parameter space of Einstein-\AE ther theory (left) and khronometric gravity (right), given observations of PSR J1141-6545~\cite{bhat}, PSR J0348+0432~\cite{2.01NS} and PSR J0737-3039~\cite{kramer-double-pulsar}. These regions account for possible variability in the NS EoS, as well as $1\sigma$ observational uncertainties in all system parameters, where we have marginalized over $(\alpha_1, \alpha_2)$ given Solar System constraints. Observe that these regions are significantly smaller than those allowed given stability/Cherenkov requirements (light, blue shaded region) and big bang nucleosynthesis constraints (dark, orange shaded region). The red dashed curves show the values of the coupling constants for which the orbital decay rate is exactly the same as in GR in the weak field/low-compactness limit~\cite{Foster:2007gr,Blas:2011zd}.}
\end{center}
\end{figure*}

%%%%%%%%%%%%%%%%%%%%%%%%%%%%%%%%%%%
{\emph{Binary Pulsar Constraints.}}--- All binary pulsar measurements of the orbital decay rate agree with the GR prediction. Thus, any deviation from this prediction must be smaller than observational uncertainties. Given the numerical sensitivities computed above, we can now evaluate the prediction of the orbital decay rate in Lorentz-violating theories and compare them to binary pulsar observations and their uncertainties.  

We here concentrate on observations of the binary pulsars PSR J1141-6545~\cite{bhat}, PSR J0348+0432~\cite{2.01NS} and PSR J0737-3039~\cite{kramer-double-pulsar}. The first two are binary systems composed of a NS and white-dwarf, either in a $0.17$ eccentricity, $4.74$-hour orbit or in a ${\cal O}(10^{-6})$ eccentricity, $2.46$-hour orbit. The last system is a double binary pulsar with $0.088$ eccentricity and $2.45$-hour orbit.

The Lorentz-violating prediction of the orbital decay rate, however, depends not only on the coupling constants of the theory, but also on the orbital period, the individual masses and the sensitivities [see~Eq~\eqref{flux-AE}]. Each of these quantities are measured to a finite accuracy that then propagates into any constraints one may wish to place. What is worse, the sensitivities and the individual masses depend on the NS EoS, which again induces a systematic uncertainty on any constraints. Given this, any given binary pulsar observation will lead to an allowed surface (instead of a line) in the $(c_{+},c_{-})$ and $(\beta,\lambda)$ parameter space, since $(\alpha_{1},\alpha_{2})$ are stringently constrained by Solar System tests and binary pulsar observations. 

One may worry that some orbital parameters, like the individual masses, are measured assuming GR is correct, and thus, it would be inconsistent to use these values to test GR. In GR, these masses are measured from the post-Keplerian parameters that depend only on the Hamiltonian of the system, e.g.~periastron precession and the Shapiro time-delay. In Lorentz-violating theories, corrections to the Hamiltonian of the system, and thus to the previous observables, are of 1PN order, ie.~${\cal{O}}(v^{2}/c^{2})$ relative to the leading order GR term. Therefore, using the values for the individual masses obtained by assuming GR is valid induces an error that is of ${\cal{O}}(v^{2}/c^{2}) \sim 10^{-6}$ for the binaries considered here.

Given $N$ binary pulsar observations, one can construct $N$, 2-dimensional allowed surfaces, all of which will be different from each other because of different system parameters and observational uncertainties. The intersection of all these surfaces yields the only allowed region in the coupling parameter space that would not be ruled out by the binary pulsar observations under consideration. Figure~\ref{fig:constraints} shows the allowed coupling parameter region given the observations of PSR J1141-6545~\cite{bhat}, PSR J0348+0432~\cite{2.01NS} and PSR J0737-3039~\cite{kramer-double-pulsar} (dark, purple shaded region). All throughout we restrict attention to values of $(\alpha_{1},\alpha_{2})$ that satisfy Solar System constraints. 

Notice that PSR J0737-3039 is very useful in constraining Lorentz-violating theories not just because of how relativistic it is, but also because both the dipolar and quadrupolar, Lorentz-violating corrections to the orbital decay rate are important for this system. This is because PSR J0737-3039 is composed of two NSs with similar masses, and thus similar sensitivities, which renders the dipolar term comparable to quadrupolar one. Thus, the orbital decay rate for this system scales with $(c_{+},c_{-})$ differently than for the other systems we considered, 
%in PSR J1141-6545~\cite{bhat} and PSR J0348+0432~\cite{2.01NS}, 
placing stronger constraints when combining all observations.  

Figure~\ref{fig:constraints} also compares the new binary pulsar constraints to other constraints in the literature. The light, blue shaded region and the dark, orange shaded one are those allowed after considering stability/Cherenkov constraints and big bang nucleosynthesis constraints respectively. We do not show cosmological constraints on Einstein-\AE ther~theory~\cite{Zuntz:2008zz} because they are comparable to the stability/Cherenkov constraints shown in the plot. Observe that binary pulsar observations push Lorentz-violating theories to a tiny region of coupling parameter space. The red, dashed curve shows the values of $(c_{+},c_{-})$ and $(\beta,\lambda)$ for which the energy flux agrees exactly with the GR prediction to leading-PN order and setting the sensitivities to zero~\cite{Foster:2007gr,Blas:2011zd}. Observe that this curve greatly underestimates the constraints that one can place with binary pulsars. The constraints on $(c_{+},c_{-})$ showed on the left panel are significantly stronger and more robust than the order-of-magnitude estimate of~\cite{Foster:2007gr}\footnote{The estimates in~\cite{Foster:2007gr} are based on a small $(c_{+},c_{-})$ approximation, leading PN-order, leading-order in the sensitivities and neglect all degeneracies, including our ignorance of the EoS.}. 

The above constraints are robust to systematic errors. The two main sources of such systematics are the neglect of the orbital eccentricity and the NS spin. The former is justified because the binary pulsar systems we considered are almost circular and eccentricity affects the dipolar radiation in the orbital decay rate at order eccentricity squared. A non-vanishing eccentricity does play an important role in other post-Keplerian parameters, like periastron-precession, which is crucial to measure the individual masses of the system. As we explained, such observables are not modified in Lorentz-violating theories to our working precision, and their measurement assuming GR is still valid. The neglect of spin is justified because the NS spin angular momentum $S_{\NS}$ observed in binary pulsars is very small: $|\vec{S}_{\NS}|/M_{\NS}^2 = {\cal{O}}(10^{-2})$~\cite{bhat,2.01NS,kramer-double-pulsar}. These systematics would modify the constraints shown here by less than $10 \%$, and would not be visible in Fig.~\ref{fig:constraints}.

%%%%%%%%%%%%%%%%%%%%%%%%%%%%%%%%%%%
{\emph{Discussion.}}--- We have derived new constraints on Lorentz-violating effects in gravity by using binary pulsar observations. We began by establishing that the modified field equations for slowly-moving NSs decouple through a tensor spherical harmonic decomposition. We then numerically integrated this decoupled differential system to obtain the NS sensitivities for a variety of EoSs, without making any weak-field assumptions. We used these sensitivities to compute binary pulsar constraints on parameters related to Lorentz violation in gravity. The new results presented here (in combination with Solar System, binary pulsar and cosmological bounds on preferred frame effects) provide the strongest constraints to date on Lorentz violation in the gravitational sector. These constraints are essential in the study of Lorentz symmetry as a fundamental property of Nature, an endeavour which may provide insights into the theory that unifies quantum mechanics and gravitational physics. Finally, we stress that a detailed discussion of our analysis and calculations can be found in Ref.~\cite{long-paper}.

%%%%%%%%%%%%%%%%%%%%%%%%%%%%%%%%%%%
{\emph{Acknowledgments}.--- We thank T.~Jacobson, S.~Gralla, T.~Tanaka and L.~Stein for providing valuable comments. We are also particularly indebted to Ted Jacobson for suggesting this
problem in the first place and for several comments.. NY acknowledges support from NSF grant PHY-1114374 and the NSF CAREER Award PHY-1250636, as well as support provided by the National Aeronautics and Space Administration from grant NNX11AI49G, under sub-award 00001944. NY and KY would like to thank the Institute d'Astrophysique de Paris and the Yukawa Institute for Theoretical Physics for their hospitality, while some of this work was being carried out. EB acknowledges support from the European Union's Seventh Framework Programme (FP7/PEOPLE-2011-CIG) through the Marie Curie Career Integration Grant GALFORMBHS PCIG11-GA-2012-321608. Some calculations used the computer algebra-systems MAPLE, in combination with the GRTENSORII package~\cite{grtensor}.

%%%%%%%%%%%%%%%%%%%%%%%%%%%%%%%%%%%%%%%%%%%%
%%%%%%%%%%%%%%%%%%%%%%%%%%%%%%%%%%%%%%%%%%%%
\bibliography{master}
\end{document}